\documentclass[a4paper,11pt]{article}

\usepackage{jinstpub} 
\usepackage{lineno}

\title{\boldmath Design, optimization and experimental characterization of RF injectors for high brightness electron beams and plasma acceleration}

\author[a,1]{V. Shpakov,\note{Corresponding author.}}
\author[a]{D. Alesini}
\author[a]{M.P. Anania}
\author[a]{M. Behtouei}
\author[a]{B. Buonomo}
\author[a]{M. Bellaveglia}
\author[a]{A. Biagioni}
\author[a]{F. Cardelli}
\author[a,b]{M. Carillo}
\author[a,b]{E. Chiadroni}
\author[a,c]{A. Cianchi}
\author[a]{G. Costa}
\author[a]{M. Del Giorno}
\author[a]{L. Faillace}
\author[a]{M. Ferrario}
\author[a]{M. del Franco}
\author[a]{G. Franzini}
\author[a,c]{M. Galletti}
\author[a,d]{L. Giannessi}
\author[a]{A. Giribono}
\author[a]{A. Liedl}
\author[a]{V. Lollo}
\author[a,b]{A. Mostacci}
\author[a]{G. Di Pirro}
\author[a]{L. Piersanti}
\author[a]{R. Pompili}
\author[a]{G. Di Raddo}
\author[a]{S. Romeo}
\author[a,b]{G. Silvi}
\author[a]{A. Stella}
\author[a]{C. Vaccarezza}
\author[a]{F. Villa}
\author[a]{A. Vannozzi}

\affiliation[a]{Laboratori Nazionali di Frascati,\\ Via Enrico Fermi 54, 00044 Frascati, Italy}
\affiliation[b]{Sapienza University,\\ Piazzale Aldo Moro 5, 00185 Rome, Italy}
\affiliation[c]{University of Rome Tor Vergata and INFN,\\ Via della Ricerca Scientifica 1, 00133 Rome, Italy}
\affiliation[d]{Elettra-Sincrotrone, \\Strada Statale 14/163.5km, 34149 Basovizza, Trieste, Italy}

\emailAdd{vladimir.shpakov@lnf.infn.it}

\abstract{In this article, we share our experience related to the new photo-injector commissioning at the SPARC\_LAB test facility.
The new photo-injector was installed into an existing machine and our goal was not only to improve the final beam parameters themselves but to improve the machine handling in day-to-day operations as well. Thus, besides the pure beam characterization, this article contains information about the improvements, that were introduced into the new photo-injector design from the machine maintenance point of view, and the benefits, that we gained by using the new technique to assemble the gun itself.}


\arxivnumber{\_\_\_\_.\_\_\_\_} 

\proceeding{\_\_\_\_\_\_\\
  \_\_\_\_\_\_\\
  \_\_\_\_\_\_}

\begin{document}
\maketitle
\flushbottom

\section{Introduction}
SPARC\_LAB is a test facility, which was initially created to develop and test various techniques for Free Electron Lasers. Later, the development of the plasma wake filed acceleration (PWFA) related technologies were added to the spectrum of our objectives \cite{ferrario2013sparc_lab}. Both tasks require a high quality electron beam, as well as a decent level of control over it \cite{filippetto2009velocity}. For these reasons, the initial choice of the photo-injector for the SPARC\_LAB facility was an S-band, 1.6 cell SLAC type electron gun \cite{alley1999design}. However, over the years, our gun has shown a number of issues. For example:
\begin{itemize}
    \item \textbf{Low quantum efficiency.} At SPARC\_LAB, we use a copper photo-cathode, which typically has quantum efficiency (QE) $\sim10^{-5}~[e/ph]$. The old gun, however, was demonstrating QE at the level of $\sim10^{-6}~[e/ph]$ \cite{pompili2021time}, and a simple cathode substitution could not solve the problem.
    \item \textbf{High dark current.} At the nominal input power, the old photo-injector was delivering up to $2~nC$ of dark current (DC). The high DC was not a problem on its own, but since our typical beam has a charge $20-200~pC$ it was creating some issues with beam charge measurements at the gun position.
    \item \textbf{High discharge rate.} Over the years, the stability of the gun has been deteriorating rather significantly, up to the point where we were observing the discharges inside of the gun every couple of minutes. Given the necessity to work with PWFA, which is inherently unstable as it is, such a discharge rate was turning into a serious issue.    
\end{itemize}
Aside from these, purely beam related issues, we have discovered several other aspects of the gun, including the mechanical ones, that we wanted to improve. Thus, the new gun was designed not only to solve the above mentioned issues, but to give us a few additional advantages from a daily operation point of view as well. 

\section{Mechanical design}
\subsection{Layout of the injector area}
\begin{figure}[htbp]
\centering
\raisebox{-0.5\height}{\includegraphics[width=0.48\textwidth]{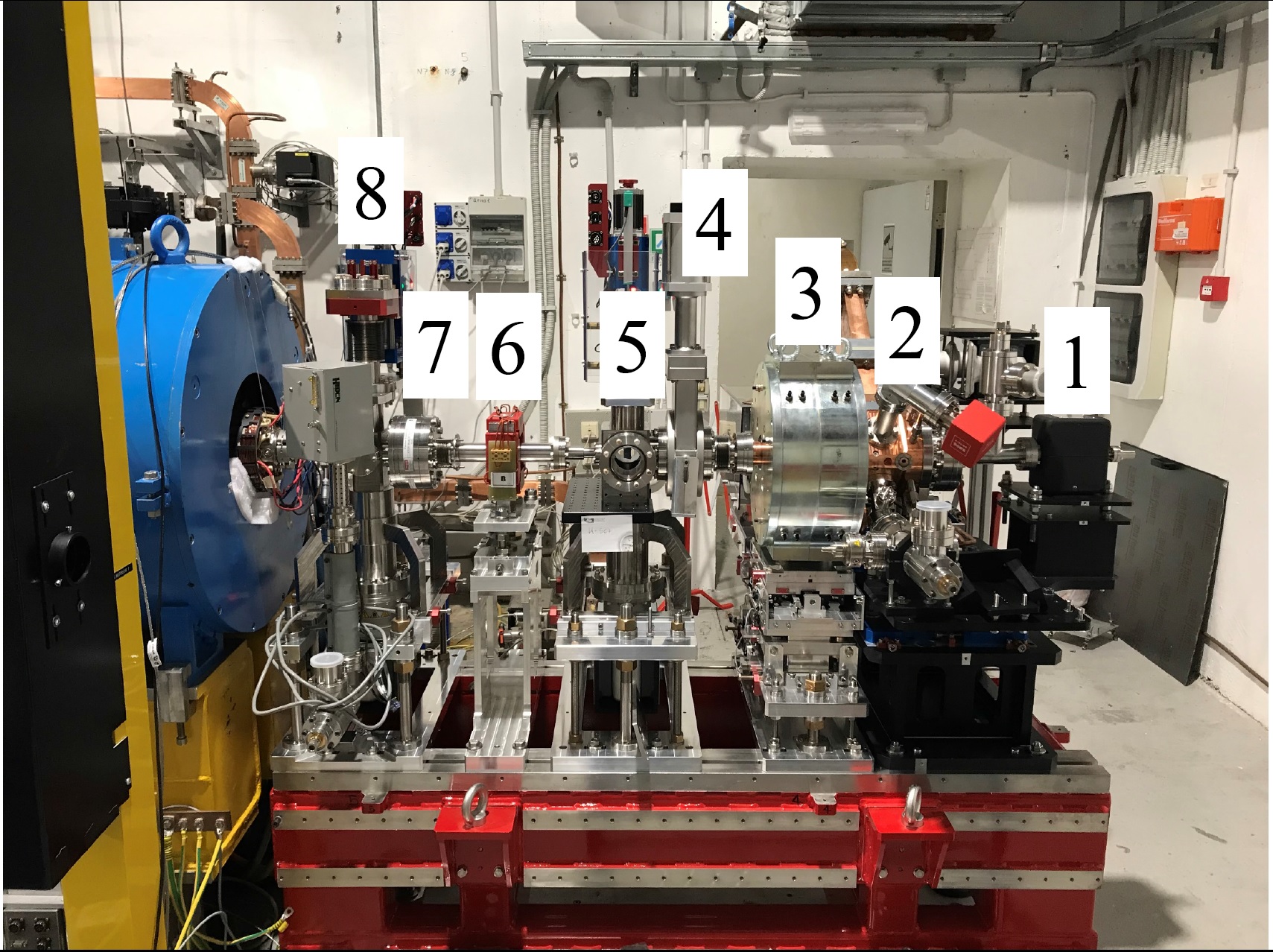}}
\raisebox{-0.5\height}{\includegraphics[width=0.48\textwidth]{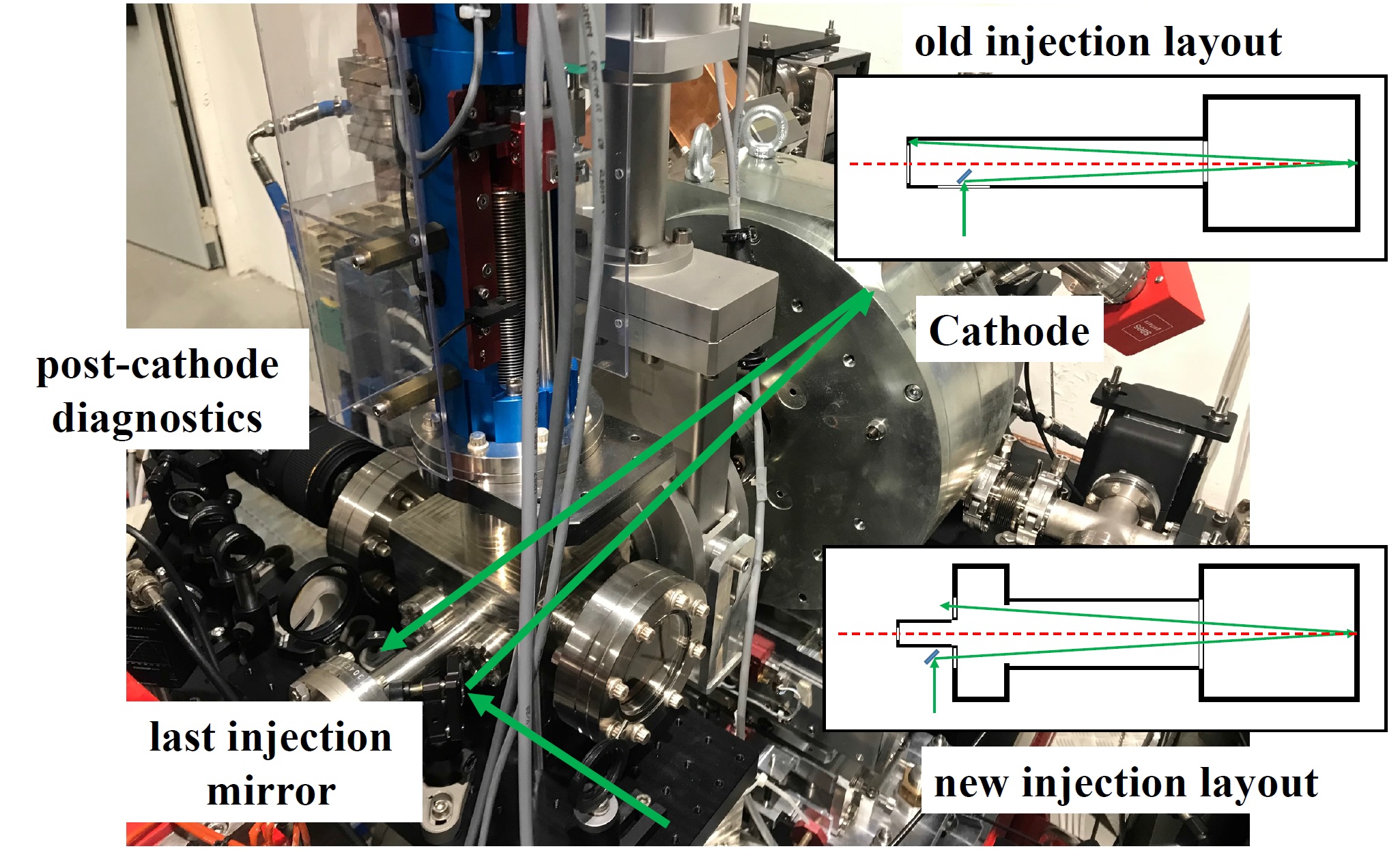}}
\caption{\label{fig1} \textbf{Left:} Fully assembled gun unit, already in SPARC bunker. The elements are: 1-vacuum pump, 2-gun, 3-solenoids, 4-valve, 5-laser injection chamber, 6-corrector, 7-BCM, 8-diagnostics flag. \textbf{Right:} New laser injection system. On the insets are the layouts of the new and old systems.}
\end{figure}

The new SPARC electron gun, which was recently adopted by the CLEAR/AWAKE project at CERN \cite{CLEARgun} as well, is a $1.6$ cell S-band photo injector (see Figure \ref{fig1}, \cite{alesini:ipac2022-MOPOMS019}). The gun itself is followed by the solenoid magnet, which consists of two coils connected with the same polarity. The solenoids are installed on 2D movable stage and their center is located at $0.2084~m$ from the face surface of the cathode (in the text, all the elements positions are reported from the face surface of the cathode). The solenoids are followed by a vacuum chamber with an off-axis laser injection windows. The gun corrector is installed at $0.82~m$ and it is used to measure the energy of the beam at the gun exit. 

The first beam current monitor (BCM) is installed right before the diagnostics flag ($1.0~m$, a combo-toroid from Bergoz \cite{bergoz}). The specific BCM was chosen in order to provide a solution for the high DC issue previously reported. The high DC on its own was not an issue, but the necessity to measure $20~pC$ beams with $2~nC$ DC was a challenge. The use of the fast current transformer (FCT) BCM could solve such a problem, since it does not read the DC. At the same time, constant DC monitoring is mandatory from radiation protection point of view, thus the integrating current transformer (ICT) was necessary as well. The chosen BCM contains both FCT and ICT in a single body, and, thus, was optimal for our purposes.    

The diagnostics flag is installed at $1.101~m$ and its YAG:Ce screen is used, among other things, for gun energy measurements. Aside from the screen, the flag is used to insert the Faraday cup into the beam line. The whole system was assembled on a single platform and was transported and installed in the accelerator bunker as a single unit. 

\subsection{Laser injection system}
To overcome the problems related to an old injection layout with a mirror inside the vacuum, namely the difficulty in laser alignment and mirror replacement, we designed a new laser injection system. The new system has two off-axis windows for the laser injection (see Figure \ref{fig1}, right), which allows us to keep all the optics in air. Such an approach automatically solves both issues mentioned above. On top of that, the second exit window allowed us to add another diagnostic to the photo-cathode laser, imaging it after the reflection from the cathode surface.
\begin{figure}[htbp]
\centering
\includegraphics[width=0.6\textwidth]{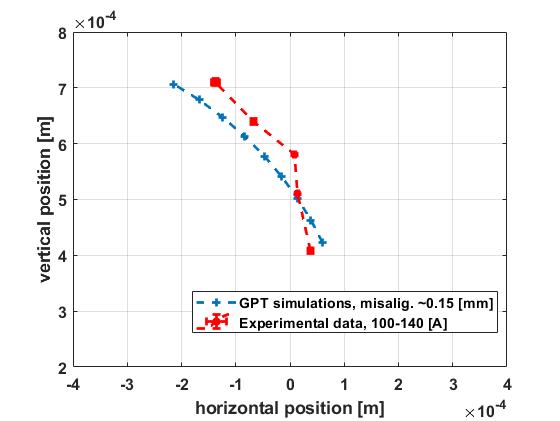}
\caption{Experimentally, measured "trajectory" of the beam center of mass during the change of the gun solenoids current from $100$ to $140~A$. In order to evaluate the absolute value of the misalignment the experimental data were compared to the GPT simulations.  
\label{GunSolAlignment}}
\end{figure}
\subsection{Gun solenoids}
\label{GunSol}
Another mechanical component that was improved over the old gun is the solenoids mounting system. The beam optimization often requires to make a scan over the solenoids current (see section \ref{Emittance}) and, if the solenoids are misaligned with respect to the gun axis, a trajectory adjustment will be necessary after each current change. On top of that, right before the main diagnostics section, at the end of the linac (see Figure \ref{SPARC}, UTLFLG02), a chamber for plasma acceleration, with vacuum impedance is placed. The impedance is only $6~mm$ in diameter, thus if the trajectory is not precise it can be a source of a charge loss.  To align the motorized solenoids we simply have found the position where the solenoids current change does not change the beam position at the first diagnostics flag (see Figure \ref{GunSolAlignment}). Such approach proved to be fast, the whole alignment process took us no more than 1 hour. According to our simulations and measurements, the residual misalignment of the gun solenoids is $\le 150~\mu m$, which proved to be sufficient for our purposes (see section \ref{Emittance}).         

\section{Gun RF design, realization and low power RF test results}
\begin{table}[htbp]
\centering
\caption{Main Parameters of the new SPARC\_LAB RF Gun (the values in parenthesis are those measured).}
\label{tab:S_band_param}
\begin{tabular}{lc}
\hline\hline
Working frequency [GHz] &2.856 (2.856) \\
$E_\text{cath}/P_\text{diss}^{1/2}$  [MV/(mMW$^{1/2}$)] & 37.5 \\
RF input power [MW] & 16\\
Cathode peak field [MV/m] & 120\\
Cathode type & copper\\
Rep. rate [Hz] & 10\\
Quality factor & 14300 (13900)\\
Filling time [ns] & 515\\
Coupling coefficient & 2 (1.98)\\
RF pulse length [$\mu s$] & 1\\
$E_\text{surf}/E_\text{cath}$ & 0.88\\
Pulsed heating [$^\circ${C}] & <30\\
Working Temperature [$^\circ${C}] & 25\\
$0-\pi$ mode frequency separation  [MHz] & 41 (41)\\
Transm. coeff. input port - antenna  [dB] & -66\\
Refl. coeff. at the input port  & -0.329\\
\hline\hline
\end{tabular}
\end{table}

\subsection{RF design}
\begin{figure}[htbp]
\centering
\raisebox{-0.5\height}{\includegraphics[width=0.45\textwidth]{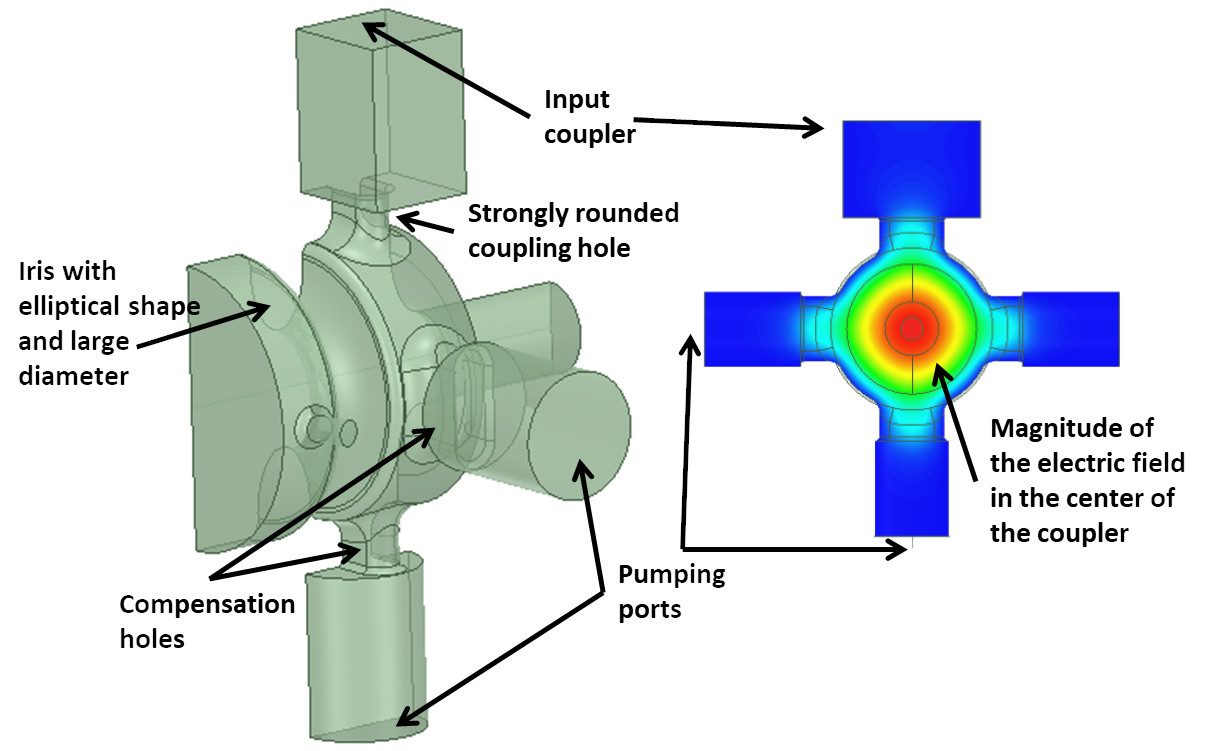}}
\raisebox{-0.5\height}{\includegraphics[width=0.45\textwidth]{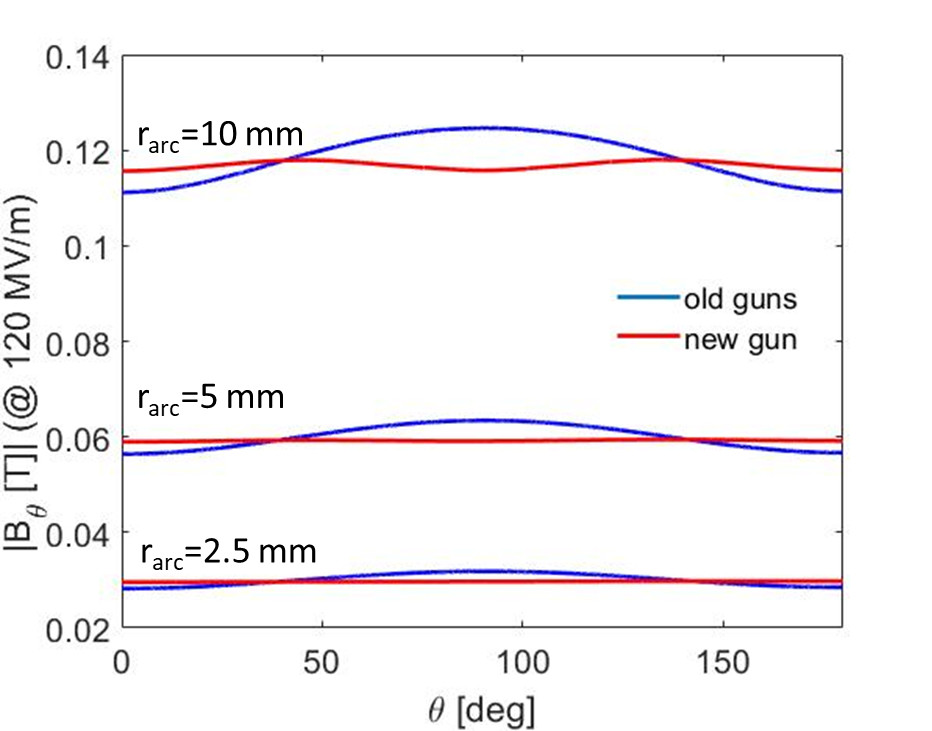}}
\caption{\label{fig3} \textbf{Left:} Electromagnetic model of the gun. \textbf{Right:} Azimuthal magnetic field on three arcs in the center of the coupling cell. }
\end{figure}

The gun design is based on a 1.6 cell RF gun fabricated with the new brazing-free technology, recently developed at the INFN-LNF \cite{AlesiniPatent}. It integrates several new features both from the electromagnetic (e.m.) and the mechanical point of view \cite{alesini:ipac2022-MOPOMS019}. The e.m. design has been accomplished using ANSYS-HFSS \cite{hfss_website} and the e.m. model is given in Figure \ref{fig3}, left. The overall design has been carried out following the same criteria already implemented in other fabricated guns \cite{alesini2015new,alesini2018design} with strongly rounded coupling holes, elliptical and larger iris diameter to simultaneously reduce the pulsed heating and the peak electric fields on the surfaces. Moreover, in this gun, we have added three holes in the coupling cell to have a full compensation of the dipole and quadrupole field components introduced by the coupling hole itself. These holes have been connected to vacuum pumps allowing a strong improvement of the vacuum pressure that, in operation, is of few $10^{-10}~mbar$ with a constant improvement. The effect of the added holes for the compensation of the quadrupole e.m. field components is well illustrated in Figure \ref{fig3}, right and Figure \ref{fig4}, left, where we have reported the azimuthal magnetic field on three different arcs in the center of the coupling cell and the related quadrupole gradient as a function of the longitudinal position respectively. The simulation results on the new gun are compared, in the figure, with those of previously fabricated guns without compensation \cite{alesini2015new,alesini2018design}. The gun has been designed with a coupling coefficient equal to 2 to allow operation with short RF pulses ($<1~\mu s$), thus reducing the breakdown rate probability (BDR) \cite{Grudiev} and the power dissipation. An elliptical profile of the iris with large aperture ($\phi=36~mm$) has been also implemented to reduce the peak electric field, to increase the $0-\pi$ mode frequency separation (thus avoiding excitation of the 0 mode with short RF pulses) and to obtain a better pumping speed on the cathode cell. The final RF gun parameters are reported in Table \ref{tab:S_band_param}. 


\subsection{Realization and low power RF test results}
\begin{figure}[htbp]
\centering
\raisebox{-0.5\height}{\includegraphics[width=0.45\textwidth]{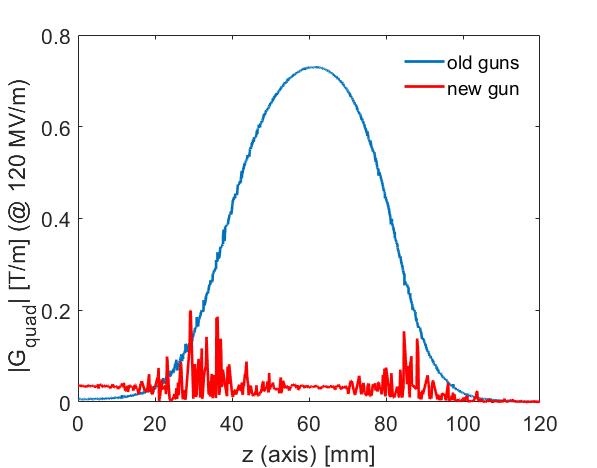}}
\raisebox{-0.5\height}{\includegraphics[width=0.45\textwidth]{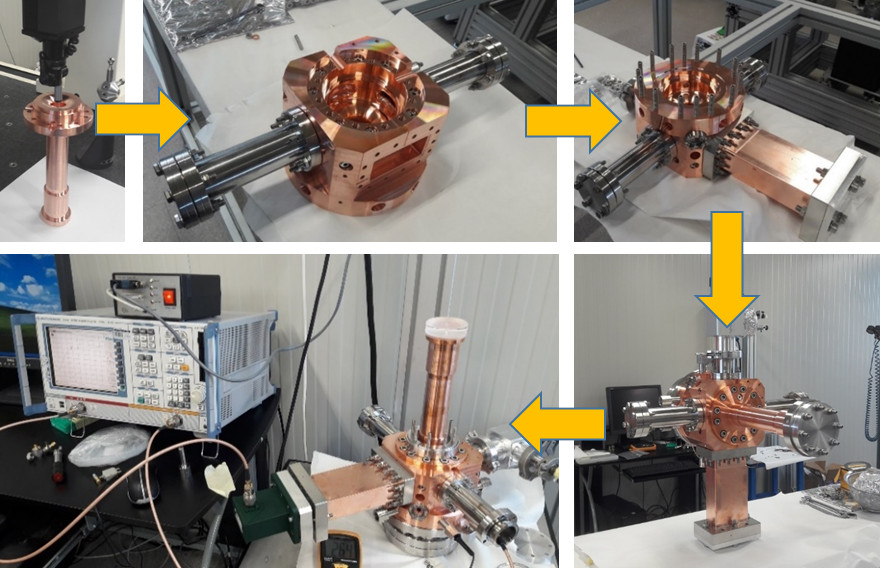}}
\caption{\label{fig4} \textbf{Left:} Equivalent quadrupole gradient as a function of the longitudinal position. \textbf{Right:} Pictures of the gun under assembly and under RF test.}
\end{figure}

The gun components were manufactured by COMEB S.r.l. \cite{comeb}. The realization technology without brazing allows to assemble the gun with special gaskets in a clean room and to proceed, after the vacuum test, directly to the RF characterization \cite{alesini2015new,alesini2018design}. Pictures of the gun under assembly and under RF test are given in Figure \ref{fig4}, right. The gun has two deformation tuners in the full cells that have been used to slightly tune the field flatness. Measurements of the reflection coefficient at the input port and transmission coefficient between the input port and the probe antenna have been performed and are reported in Table \ref{tab:S_band_param} (the measurements have been performed under the nitrogen at $22^{o}~C$). Electric field measurements have been performed using the bead drop technique with a small metallic sphere with a diameter of $\oslash 2~mm $ and have demonstrated flatness of the field on par with the theoretical calculations. 




\section{Gun conditioning and beam parameters measurements}

\subsection{Conditioning process}

\begin{figure}[htbp]
\includegraphics[width=0.98\textwidth]{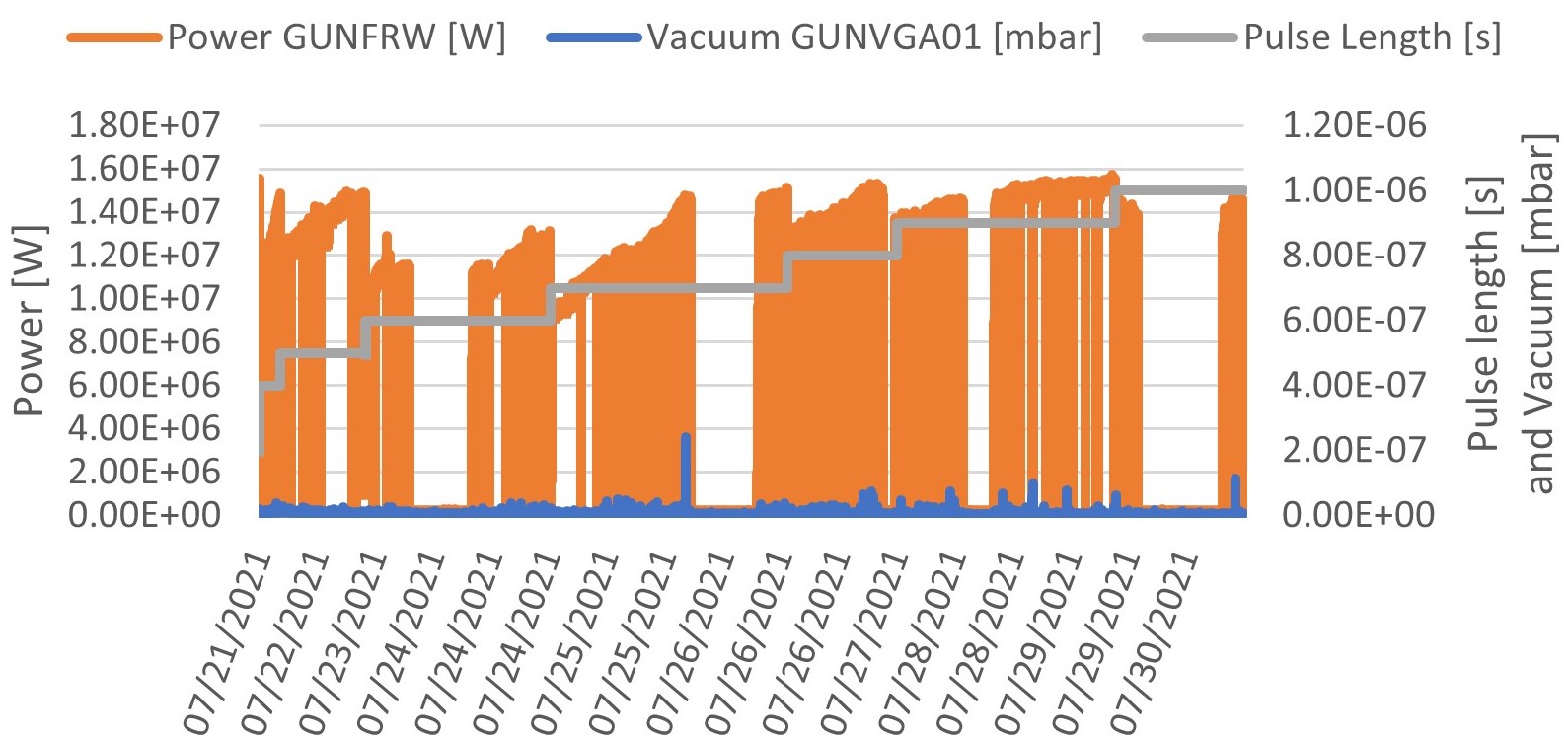}
\caption{Plot of the RF gun conditioning data. We reported the behavior of incident power, pulse length and vacuum level.  
\label{fig5}}
\end{figure}

After the low level characterization in the RF laboratory, the gun has been tested at high power, reaching the nominal specifications ($16~MW$ incident power, $120~MV/m$ gradient) in a remarkably short time (less than 10 days at $10~Hz$). The conditioning data archive is reported in Figure \ref{fig5} where the measured vacuum pressure, the RF input power and the pulse length behaviors are plotted.

The conditioning has been carried out using a semi-automatic algorithm, that takes into account the vacuum level for the interlock generation. Indeed, in case of an RF breakdown and, consequently, vacuum pressure exceeding a given threshold (typically $5-6\times10^{-8}~mbar$), the system automatically stopped the RF power, waited for the vacuum recovery (typically $2-3\times10^{-8}~mbar$), and then restarted again with a power reduction of $0.1~dB$. During the conditioning the repetition rate has been kept constant at $10~Hz$, and the pulse length has been progressively increased up to the nominal value of $1~\mu s$. For each pulse length step, the incident input power level has been increased up to $16~MW$. The power increase always required a manual operation.

During the injector conditioning (performed on a parallel beam line) the solenoids were off. During the 10 days of conditioning, $\approx5\times10^6$ pulses have been delivered to the gun, with a total number of detected discharges lower than $10^3$. The final breakdown rate (BDR) has been measured in a 24-hour operation, and it has been estimated to be below $10^{-6}$ breakdown per pulse.

After the conditioning process, the gun has been vented in dry nitrogen and moved in its final position. It was then feeded again with the RF pulse and brought back to the nominal power, also switching on the focusing solenoid. The effects of the position change and of the solenoid magnetic field on the BDR were completely negligible and the gun recovered the final performances without any considerable effect.

\subsection{Gun energy and QE}
\begin{figure}[htbp]
\centering
\raisebox{-0.5\height}{\includegraphics[width=0.48\textwidth]{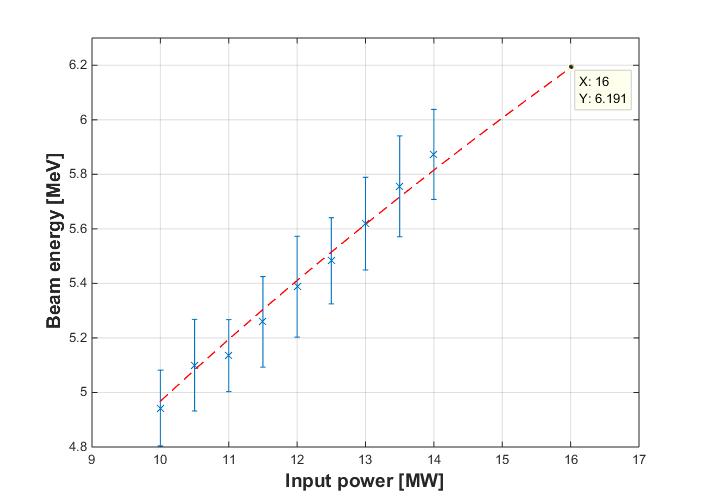}}
\raisebox{-0.5\height}{\includegraphics[width=.48\textwidth]{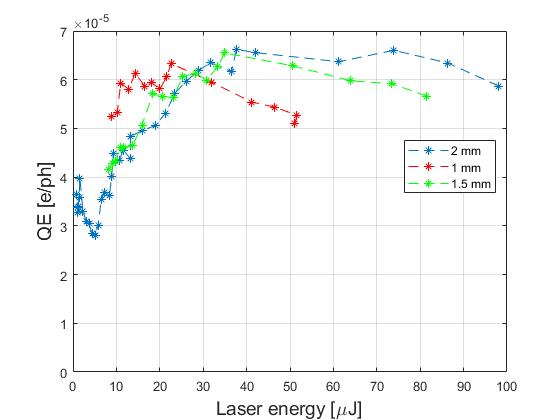}}
\caption{\textbf{Left:} The energy of the beam at the exit from the gun (blue crosses) as a function of the input power. The red dashed line is the simulations made by GPT.
\textbf{Right:} Quantum efficiency of the SPARC gun (January 2022) for the different diameters of the laser spot at the cathode. The drop of the QE at low charge/laser energy is related to our inability to measure properly the energy of the laser pulse at such levels.
\label{fig6}}
\end{figure}

The energy of the beam at the exit from the gun was measured using the first corrector (see Fig\ref{fig1}, left). The energy of the beam was always measured at $40^\circ$ launch phase from the "zero crossing", which corresponds to a maximum achievable energy for a given input power, according to our simulations. The experimentally measured beam energy as a function of the input power is shown in Figure \ref{fig6}, left. The measured beam energy at the gun exit was used in order to calibrate the the parameters for GPT simulations, which allowed us to estimate the achieved peak field in the gun. The maximum energy, $\approx5.8~MeV$, was measured at the $14~MW$ input power. According to the GPT simulations, such energy corresponds to the peak field $\approx111.9~MV/m$. It is worth to mention, that according to the theoretical estimation the maximum field achievable in the gun is suppose to be $120~MV/m$, which correspond to the $\sim16~MW$ input power and $\approx6.2~MeV$ beam energy. However, despite of the fact that the gun has demonstrated rather high level of the vacuum stability (see the following subsection), we could not push the power to such levels due to the limitations of the waveguide system. 

After establishing our maximum energy working point, we used it for QE measurements. During the QE measurements the UV laser had a duration $\tau_{rms}\approx130-150~fs$ rms and the flat-top spot at the cathode surface $\oslash 2~mm$. The results of the quantum efficiency measurements are depicted in Figure \ref{fig6}, right. The measured QE for the gun was at the level of $\sim6.0\times10^{-5} e./ph.$. The routine operation at SPARC requires beams with the charge in range $20-200~pC$, but during the QE measurements the gun was pushed to $\sim1200~pC$ beam charge (rightmost point at Figure \ref{fig6}, right, $\approx98~\mu J$ laser energy).

The QE measurements were conducted also for the $1.0~mm$ and $1.5~mm$ laser spot sizes at the cathode. In both cases the maximum QE was at the same level $\sim6.0\times10^{-5} e./ph.$. However, for such small laser diameters, we observed the effects of the electron emission saturation (image charge, space charge effects). For $1~mm$ spot size, the feature of the saturation regime, a substantial drop in QE, was observed already at $600~pC$ ($\approx50~\mu J$ laser energy). 
Overall, the QE of cathode falls within our expectation limits for the copper. It is necessary to highlight, that at SPARC\_LAB we use quite short laser pulses (hundreds$~fs$ regime) due to the requirements of our PWFA related experiments \cite{pompili2016beam,chiadroni2017beam}, which leads to a much faster emission saturation. Using longer laser pulses, $\sim5~ps$ for example, not only will push further the saturation thresholds, but it will allow to extract a higher charge as well.     
\subsection{Dark current and gun stability}
\begin{figure}[htbp]
\centering
    \includegraphics[width=.6\textwidth]{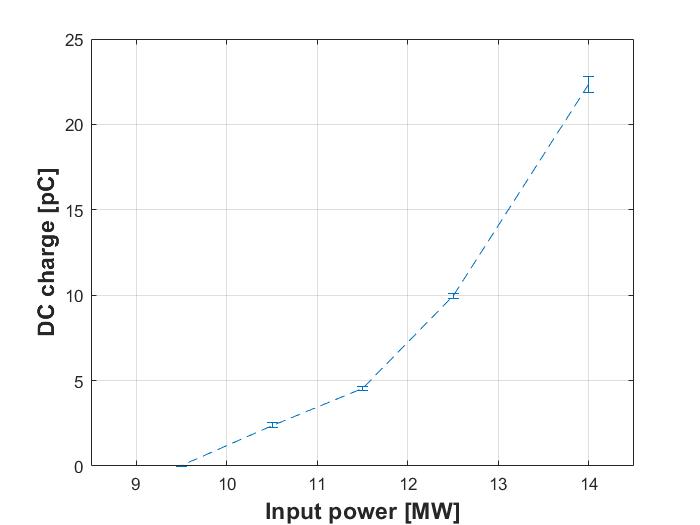}
    \caption{Dark current measurements as a function of the gun input power.}
  \label{DC}
\end{figure}

As it was indicated in the introduction, one of the issues that we wanted to solve with the new gun was the high dark current, up to $\sim2~nC$ for the old gun at the nominal input power. The dark current for the new gun is 2 orders of magnitude lower (see Figure \ref{DC}). Even at the nominal power of $14~MW$, the DC falls within the $20-25~pC$ range. On top of that, if the input power decreases, the DC rapidly falls down and at the level of $10~MW$ it is below of the $pC$ range.

\begin{figure}[htbp]
\centering
\raisebox{-0.5\height}{\includegraphics[width=0.48\textwidth]{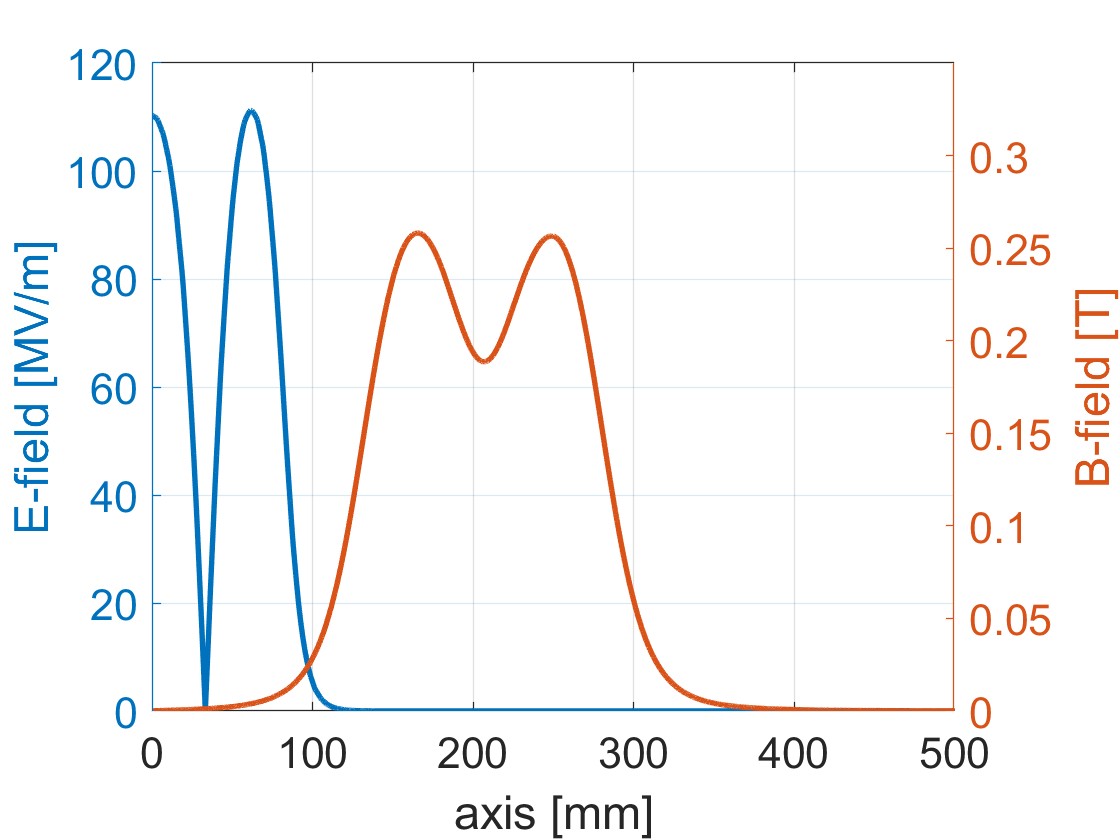}}
\raisebox{-0.5\height}{\includegraphics[width=0.48\textwidth]{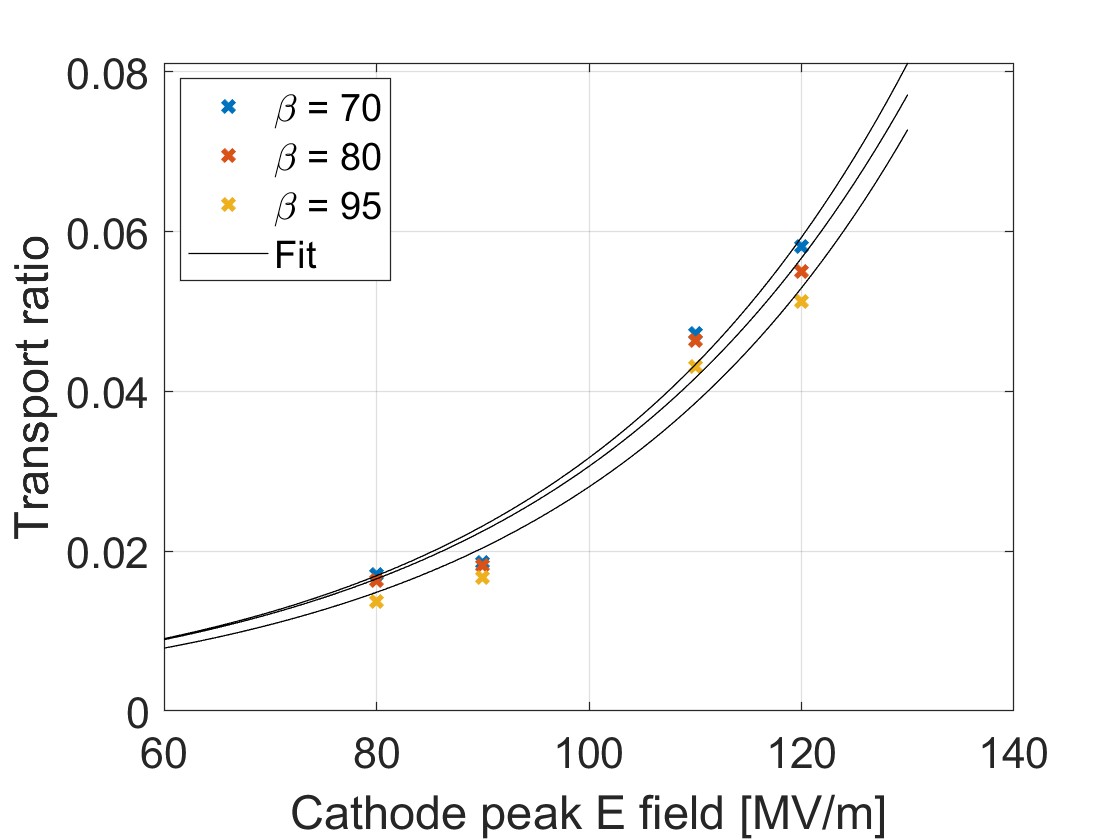}}
\caption{\textbf{Left:} RF longitudinal electric field and solenoids magnetic field along the gun axis.\textbf{Right:} Transport ratio between the charge transported and emitted in a RF period as function of the cathode maximum peak field for different the values of the emission coefficient
\label{fig8}}
\end{figure}

Before the gun commissioning, to estimate the electrons field emitted from the cathode both a theoretical approach based on the Fowler-Nordheim (FN) emission model and a numerical approach using the 3D CST Particle-in-cell code have been used and compared with each other. As source of particles, the entire cathode surface has been considered and an emitter density factor has been introduced to scale the effective area of the emitters. The emitters area $A_e$ and the FN emission coefficient $\beta$ characterize the entire emission process and depend mainly on the cleanliness and roughness of the emitting surface. Those are unpredictable, so as a starting point for these two parameters, to make a preliminary DC estimation, we took $\beta=70$ and $A_e=0.01~\mu m^2$, evaluated at PSI for the SwissFEL Gun \cite{bettoni2018low}. The PIC simulations are based on the technique described in \cite{cardelli:ipac2022-mopoms020} and include the solenoid model, which allows to estimate the energy spectrum and the DC beam transport along the photo-injector up to the diagnostics chamber. The gun RF field and the magnetic field of the solenoid have been previously simulated adopting this geometry and then imported into the PIC simulations. An example of the fields along the gun axis, calculated by CST for a cathode peak field of $111.9~MV/m$ are reported in Figure \ref{fig8}, left.

The transport along the injector has been investigated performing different simulations by varying the values of the cathode peak field and of the emission coefficient $\beta$. Figure \ref{fig8}, right reports the transport factor obtained by the ratio between the charge transported and emitted in a RF period. As expected, it shows a negligible dependence with respect to the emission coefficient variation.

By means of the fit of the transport and the analytically evaluated emitted charge, it is possible to fit the transported charge and obtain, given the RF pulse field in the gun, the total DC charge emitted in an entire pulse. Thus, from the simulation with $\beta = 70$ and $A_e = 0.01~\mu m^2$, the DC charge transported during the RF pulse was $3~pC$, lower than the one measured during the commissioning. This implies that the mechanical finishing of the cathode is slightly worse than the one evaluated for the SwissFEL cathode, considered as a reference, but this is still within an acceptable range given the low DC value measured. The comparison between the measured values and the simulated one allows to estimate the effective enhancement factor $\beta$ in the FN field emission expression for our gun. According to our estimations the $\beta \sim 89$, assuming constant the effective area of the emitters ($A_e = 0.01~\mu m^2$), corresponds to $\sim 22~pC$ DC at the diagnostics chamber. 

\begin{figure}[h]
\raisebox{-0.5\height}{\includegraphics[width=.48\textwidth]{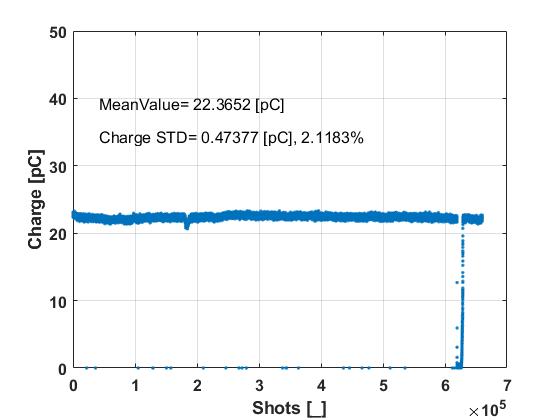}}
\raisebox{-0.5\height}{\includegraphics[width=.48\textwidth]{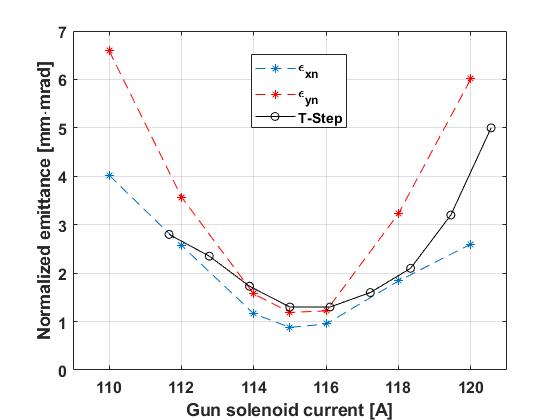}}
\caption{\textbf{Left:} 24 hour cycle DC measurement. The measurement has demonstrated no discharges at all. The drop at the end of the cycle was caused by a lack of communication between the front end and the control system. \textbf{Right:} Example of the gun solenoid scan for emittance optimization for the typical $200~pC$ SPARC beam. Dashed lines are the experimental data and the solid line is a theoretical curve made with TStep code.}
  \label{fig9}
\end{figure}

During the gun commissioning the DC was also used to evaluate the gun discharge rate. During the discharge inside of the gun, the DC experiences a temporary spike. In Figure \ref{fig9}, left is depicted a 24 hour long measurement cycle, during which we could not record even a single gun discharge event. The input power was set at the nominal value of $14~MW$ during the whole measurement and the solenoid was under the normal working current. It is worth to mention, that during the first several month we have observed a few gun discharges, however, they are so rare that it is simply not even a daily event. Besides the rather low discharge rate, these measurements also have demonstrated a high level of DC stability.  

\section{Emittance}\label{Emittance}
Unlike the parameters described above (QE, gun energy, etc...), the beam emittance is not purely defined by the electron gun. A number of parameters related to the linac and the beam itself (beam charge, for example) will play a significant role in shaping the emittance. Thus, in order to demonstrate the "emittance capability" of the new gun, here we use our typical beam - a driver for our PWFA experiments \cite{pompili2020energy,pompili2022free}. 
\subsection{Emittance of a typical beam}
\begin{figure*}
  \includegraphics[width=\textwidth]{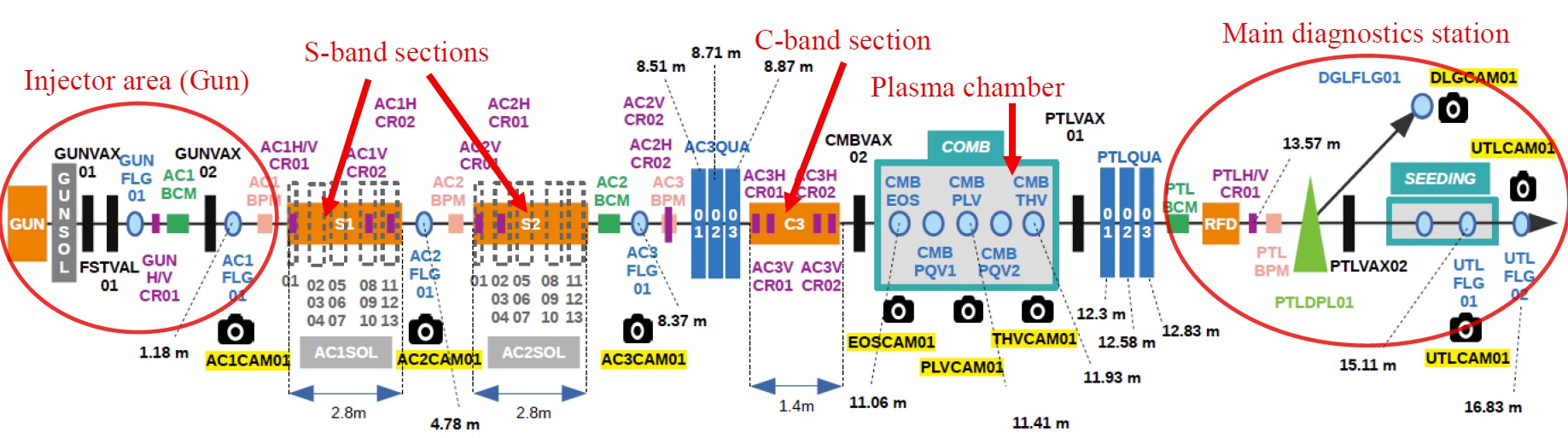}
  \caption{SPARC layout. All distances are indicated as measured from the cathode surface.}
  \label{SPARC}
\end{figure*}

At SPARC\_LAB, we use relatively short laser pulses to 
extract the electrons from the cathode, thus, for the emittance measurements was used a laser pulse with rms duration $\sim150~fs$, with flap-top profile on the cathode surface and diameter of $1~mm$. The duration of the beam at the main diagnostics station was $\sim1.2-1.4~ps$, no RF compression was used. The final energy was $140~MeV$ and the beam charge was $200~pC$. Another peculiarity of the beam transport at SPARC\_LAB, above mentioned, is the plasma chamber (see Figure \ref{SPARC}). The plasma chamber has a few vacuum impedances inside, with $6~mm$ aperture, thus, in order to go through the chamber without charge losses, we usually use the section solenoids to focus the beam at the plasma chamber entrance down to $\sim200\mu m$ rms transverse size.       

The beam emittance optimization at SPARC\_LAB includes the gun solenoid scan and the example of such optimization is depicted in Figure \ref{fig9}, right. Each point in Figure \ref{fig9}, right is a quadrupole scan made by PTLQUAD01 quadrupole and measured at UTLFLG02 diagnostics screen (see Figure \ref{SPARC}). Thus, a typical $200~pC$ beam at SPARC photo-injector has a normalized emittance near $1~mm\cdot mrad$. The experimental results were compared with simulations made with a particle tracker TStep \cite{TStep2}. 
Another item, worth to mention here, is the alignment procedure described in the section \ref{GunSol}, since the solenoid scan depicted in Figure \ref{fig9},right is one of the main reasons to perform such alignment. The conclusion regarding sufficiency of the alignment procedure was made from the fact that the typical solenoid scan, just like in Figure \ref{fig9}, right or in Figure \ref{EmittanceDecember2021}, does not require any trajectory adjustments for different values of the solenoid current. 

\subsection{Gun quadrupole} 
\begin{figure}[h]
\raisebox{-0.5\height}{\includegraphics[width=.48\textwidth]{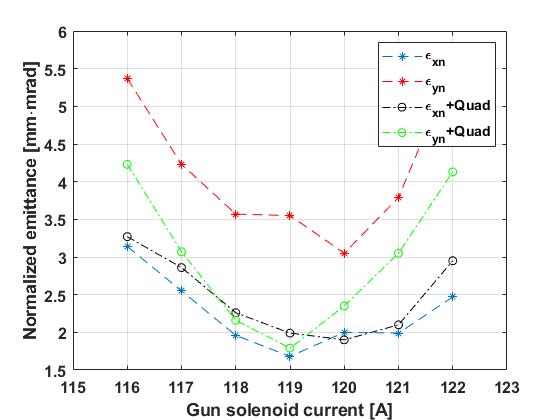}}
\raisebox{-0.5\height}{\includegraphics[width=.23\textwidth]{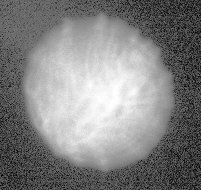}}
\raisebox{-0.5\height}{\includegraphics[width=.23\textwidth]{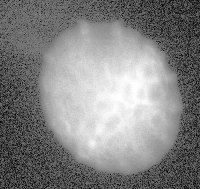}}
    \caption{\textbf{Left:} Example of the emittance correction done by the built-in gun quadruple. \textbf{Right:} The effect of the gun quadrupole on the beam, as it seen right at the exit from the gun: quadruple OFF and quadrupole ON correspondingly.}
  \label{EmittanceDecember2021}
\end{figure}

The new gun, aside from motorized solenoid, includes an additional quadrupole, which is intended for emittance manipulation. In order to demonstrate the capability of that quadrupole, we have created a highly asymmetric beam in terms of the emittance (see Figure \ref{EmittanceDecember2021}, blue and red stars). Such effect can be achieved by mismatching the solenoids around the accelerating sections. Specifically, we have created the beam with $1.69~m\cdot mrad$ horizontal and $3.05~mm\cdot mrad$ vertical emittance. Such emittance asymmetry was successfully corrected with the help of the new gun quadrupole (see Figure \ref{EmittanceDecember2021}, black and green circles). Naturally, such a correction is achieved by creating an asymmetry in terms of the beam transverse profile right at the exit from the gun, which can be seen in Figure \ref{EmittanceDecember2021} as well. 

In this particular case, the emittance asymmetry was created artificially, and in our daily operation, with proper beam transport (see Figure \ref{fig9}, right), the gun quadrupole is not used. Nevertheless, we have demonstrated that such quadrupole can be used in order to compensate emittance asymmetry or to create one, if necessary.  

\subsection{Gun intrinsic emittance}

\begin{figure}
\centering
    \includegraphics[width=.6\textwidth]{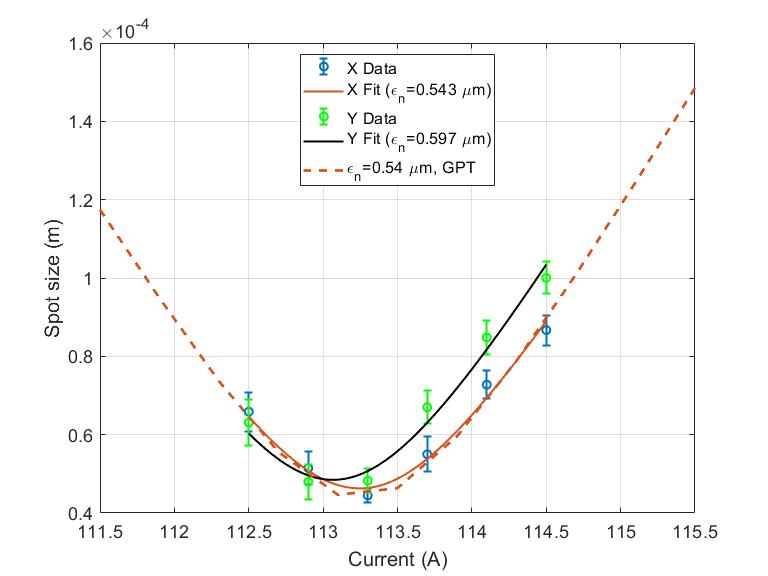}
    \caption{Gun intrinsic emittance, measured right at the exit from the gun. The scan performed using the gun solenoids. The beam charge here is only $0.3~pC$, with the beam energy $5.82~MeV$. The diameter of the laser spot at the cathode is $2~mm$}
  \label{GunEmittance}
\end{figure}

The emittance measured at the end of the linac depends on a wide range of parameters and it can be hardly associated with the gun only. However, at SPARC\_LAB there is the possibility to measure the emittance right at the exit of the gun using the gun solenoids to focus the beam. The procedure to measure such an intrinsic emittance is identical to the classical quadrupole scan with one fundamental difference. The electron beam at the exit from the gun has the energy of only $\sim5.8~MeV$ and simply not rigid enough. The high charge beam, while passing through the waist of the scan (see Figure \ref{GunEmittance}), will experience significant space charge effects, making the measurement not valid. Thus the gun intrinsic emittance is always measured for rather low charge beams to avoid any possible space charge effects.

At SPARC\_LAB, the intrinsic emittance was measured at the first diagnostics flag (see Figure \ref{SPARC}, AC1CAM01). The distance between the gun solenoid and the screen was $\sim1~m$, the beam energy was $5.82~MeV$. For our parameters, according to the simulations \cite{gpt_web}, the beam charge should be lowered to $0.5~pC$ or less to avoid the space charge effects during the scan. The charge of the beam was lowered down to $0.3~pC$. For the $2~mm$ diameter, flat-top laser spot at the cathode, the intrinsic emittance resulted in $\sim0.55~\mu m$, which is in accordance with our simulations (see Figure \ref{GunEmittance}). The decrease of the laser spot size down to $1~mm$, with the charge remaining at $0.3~pC$, has led to the reduction of the emittance to $0.35~\mu m$, which was also expected from the simulations.

\section{Conclusion}
The main reason behind the substitution of the gun at SPARC\_LAB was the necessity to solve several issues related to the old gun, indicated in the introduction. The new gun has demonstrated QE at the level of $\sim10^{-5}~[e/ph]$ which was our initial desire and is typical for the copper cathodes. The DC at the nominal input power has dropped by two orders of magnitude, from $\sim2~nC$ to $\sim22~pC$, which is also more then satisfactory.

Thanks to the use of the new brazing-less technique to assemble the gun, we managed to achieve two major improvements. First, the discharge rate dropped down to a practically zero level. Our measurements have demonstrated a capability of such gun to work at nominal power $24+$ hours without any discharges. Second, a major improvement is the conditioning time, which dropped significantly down to $\leq2$~weeks.

Since, effectively, we were substituting a 1.6 cell, S-band gun with another 1.6-cell, S-band gun, we did not expected any major improvements in the raw beam quality as our measurements have demonstrated. Aside from the fact that, thanks to the restored QE, we are now capable to produce beams with much higher charge, the upgraded photo-injector is capable to provide $200~pC, 1.4~ps$ long beam with the normalized emittance at $1~mm\cdot mrad$ and the intrinsic emittance at $\sim0.5~mm\cdot mrad$, which is on par with other injectors with similar designs \cite{di2013emittance,palmer1997initial,zheng2016development,divall2015intrinsic}.

With the new gun, we have introduced a number of mechanical improvements. The new laser injection system without in-vacuum mirrors, appeared to be much more reliable from the laser alignment point of view compared to the old injection system. The motorized solenoids allowed us to perform a simple, but rather efficient beam based alignment, which made our daily beam operation much easier. The new in-build gun quadrupole has demonstrated its efficiency in the emittance manipulation. Finally, the idea to build a gun with all the surrounding equipment (e.g. vacuum pumps, optical elements, diagnostics, etc.) as a single unit proved to be advantageous, since the whole system is assembled in the laboratory and simply transported to the bunker for installation as a single block.   

\acknowledgments

The SPARC group would like to express its gratitude to Paolo Chimenti and Antonio Battisti for the help with the gun handling and installation.
This work was partially funded by SABINA project, co-funded by Regione Lazio with the "Research infrastructures" public call, within PORFESR 2014-2020 program.


\bibliography{biblio}

\begin{thebibliography}{25}%
\makeatletter
\providecommand \@ifxundefined [1]{%
 \@ifx{#1\undefined}
}%
\providecommand \@ifnum [1]{%
 \ifnum #1\expandafter \@firstoftwo
 \else \expandafter \@secondoftwo
 \fi
}%
\providecommand \@ifx [1]{%
 \ifx #1\expandafter \@firstoftwo
 \else \expandafter \@secondoftwo
 \fi
}%
\providecommand \natexlab [1]{#1}%
\providecommand \enquote  [1]{``#1''}%
\providecommand \bibnamefont  [1]{#1}%
\providecommand \bibfnamefont [1]{#1}%
\providecommand \citenamefont [1]{#1}%
\providecommand \href@noop [0]{\@secondoftwo}%
\providecommand \href [0]{\begingroup \@sanitize@url \@href}%
\providecommand \@href[1]{\@@startlink{#1}\@@href}%
\providecommand \@@href[1]{\endgroup#1\@@endlink}%
\providecommand \@sanitize@url [0]{\catcode `\\12\catcode `\$12\catcode
  `\&12\catcode `\#12\catcode `\^12\catcode `\_12\catcode `\%12\relax}%
\providecommand \@@startlink[1]{}%
\providecommand \@@endlink[0]{}%
\providecommand \url  [0]{\begingroup\@sanitize@url \@url }%
\providecommand \@url [1]{\endgroup\@href {#1}{\urlprefix }}%
\providecommand \urlprefix  [0]{URL }%
\providecommand \Eprint [0]{\href }%
\providecommand \doibase [0]{http://dx.doi.org/}%
\providecommand \selectlanguage [0]{\@gobble}%
\providecommand \bibinfo  [0]{\@secondoftwo}%
\providecommand \bibfield  [0]{\@secondoftwo}%
\providecommand \translation [1]{[#1]}%
\providecommand \BibitemOpen [0]{}%
\providecommand \bibitemStop [0]{}%
\providecommand \bibitemNoStop [0]{.\EOS\space}%
\providecommand \EOS [0]{\spacefactor3000\relax}%
\providecommand \BibitemShut  [1]{\csname bibitem#1\endcsname}%
\let\auto@bib@innerbib\@empty
\bibitem [{\citenamefont {Ferrario}\ \emph {et~al.}(2013)\citenamefont
  {Ferrario}, \citenamefont {Alesini}, \citenamefont {Anania}, \citenamefont
  {Bacci}, \citenamefont {Bellaveglia}, \citenamefont {Bogdanov}, \citenamefont
  {Boni}, \citenamefont {Castellano}, \citenamefont {Chiadroni}, \citenamefont
  {Cianchi} \emph {et~al.}}]{ferrario2013sparc_lab}%
  \BibitemOpen
  \bibfield  {author} {\bibinfo {author} {\bibfnamefont {M.}~\bibnamefont
  {Ferrario}}, \bibinfo {author} {\bibfnamefont {D.}~\bibnamefont {Alesini}},
  \bibinfo {author} {\bibfnamefont {M.}~\bibnamefont {Anania}}, \bibinfo
  {author} {\bibfnamefont {A.}~\bibnamefont {Bacci}}, \bibinfo {author}
  {\bibfnamefont {M.}~\bibnamefont {Bellaveglia}}, \bibinfo {author}
  {\bibfnamefont {O.}~\bibnamefont {Bogdanov}}, \bibinfo {author}
  {\bibfnamefont {R.}~\bibnamefont {Boni}}, \bibinfo {author} {\bibfnamefont
  {M.}~\bibnamefont {Castellano}}, \bibinfo {author} {\bibfnamefont
  {E.}~\bibnamefont {Chiadroni}}, \bibinfo {author} {\bibfnamefont
  {A.}~\bibnamefont {Cianchi}},  \emph {et~al.},\ }\href {\doibase
  10.1016/j.nimb.2013.03.049} {\bibfield  {journal} {\bibinfo  {journal}
  {Nuclear Instruments and Methods B}\ }\textbf {\bibinfo {volume} {309}},\
  \bibinfo {pages} {183} (\bibinfo {year} {2013})}\BibitemShut {NoStop}%
\bibitem [{\citenamefont {Filippetto}\ \emph {et~al.}(2009)\citenamefont
  {Filippetto}, \citenamefont {Alesini}, \citenamefont {Bellaveglia},
  \citenamefont {Boni}, \citenamefont {Boscolo}, \citenamefont {Castellano},
  \citenamefont {Chiadroni}, \citenamefont {Cultrera}, \citenamefont
  {Di~Pirro}, \citenamefont {Ferrario} \emph
  {et~al.}}]{filippetto2009velocity}%
  \BibitemOpen
  \bibfield  {author} {\bibinfo {author} {\bibfnamefont {D.}~\bibnamefont
  {Filippetto}}, \bibinfo {author} {\bibfnamefont {D.}~\bibnamefont {Alesini}},
  \bibinfo {author} {\bibfnamefont {M.}~\bibnamefont {Bellaveglia}}, \bibinfo
  {author} {\bibfnamefont {R.}~\bibnamefont {Boni}}, \bibinfo {author}
  {\bibfnamefont {M.}~\bibnamefont {Boscolo}}, \bibinfo {author} {\bibfnamefont
  {M.}~\bibnamefont {Castellano}}, \bibinfo {author} {\bibfnamefont
  {E.}~\bibnamefont {Chiadroni}}, \bibinfo {author} {\bibfnamefont
  {L.}~\bibnamefont {Cultrera}}, \bibinfo {author} {\bibfnamefont
  {G.}~\bibnamefont {Di~Pirro}}, \bibinfo {author} {\bibfnamefont
  {M.}~\bibnamefont {Ferrario}},  \emph {et~al.},\ }in\ \href@noop {} {\emph
  {\bibinfo {booktitle} {Proc. of the FEL 2009 conference, Liverpool, UK}}}\
  (\bibinfo {year} {2009})\BibitemShut {NoStop}%
\bibitem [{\citenamefont {Alley}\ \emph {et~al.}(1999)\citenamefont {Alley},
  \citenamefont {Bharadwaj}, \citenamefont {Clendenin}, \citenamefont {Emma},
  \citenamefont {Fisher}, \citenamefont {Frisch}, \citenamefont {Kotseroglou},
  \citenamefont {Miller}, \citenamefont {Palmer}, \citenamefont {Schmerge}
  \emph {et~al.}}]{alley1999design}%
  \BibitemOpen
  \bibfield  {author} {\bibinfo {author} {\bibfnamefont {R.}~\bibnamefont
  {Alley}}, \bibinfo {author} {\bibfnamefont {V.}~\bibnamefont {Bharadwaj}},
  \bibinfo {author} {\bibfnamefont {J.}~\bibnamefont {Clendenin}}, \bibinfo
  {author} {\bibfnamefont {P.}~\bibnamefont {Emma}}, \bibinfo {author}
  {\bibfnamefont {A.}~\bibnamefont {Fisher}}, \bibinfo {author} {\bibfnamefont
  {J.}~\bibnamefont {Frisch}}, \bibinfo {author} {\bibfnamefont
  {T.}~\bibnamefont {Kotseroglou}}, \bibinfo {author} {\bibfnamefont
  {R.}~\bibnamefont {Miller}}, \bibinfo {author} {\bibfnamefont
  {D.}~\bibnamefont {Palmer}}, \bibinfo {author} {\bibfnamefont
  {J.}~\bibnamefont {Schmerge}},  \emph {et~al.},\ }\href@noop {} {\bibfield
  {journal} {\bibinfo  {journal} {Nuclear Instruments and Methods in Physics
  Research Section A: Accelerators, Spectrometers, Detectors and Associated
  Equipment}\ }\textbf {\bibinfo {volume} {429}},\ \bibinfo {pages} {324}
  (\bibinfo {year} {1999})}\BibitemShut {NoStop}%
\bibitem [{\citenamefont {Pompili}\ \emph {et~al.}(2021)\citenamefont
  {Pompili}, \citenamefont {Chiadroni}, \citenamefont {Cianchi}, \citenamefont
  {Curcio}, \citenamefont {Del~Dotto}, \citenamefont {Ferrario}, \citenamefont
  {Galletti}, \citenamefont {Romeo}, \citenamefont {Scifo}, \citenamefont
  {Shpakov} \emph {et~al.}}]{pompili2021time}%
  \BibitemOpen
  \bibfield  {author} {\bibinfo {author} {\bibfnamefont {R.}~\bibnamefont
  {Pompili}}, \bibinfo {author} {\bibfnamefont {E.}~\bibnamefont {Chiadroni}},
  \bibinfo {author} {\bibfnamefont {A.}~\bibnamefont {Cianchi}}, \bibinfo
  {author} {\bibfnamefont {A.}~\bibnamefont {Curcio}}, \bibinfo {author}
  {\bibfnamefont {A.}~\bibnamefont {Del~Dotto}}, \bibinfo {author}
  {\bibfnamefont {M.}~\bibnamefont {Ferrario}}, \bibinfo {author}
  {\bibfnamefont {M.}~\bibnamefont {Galletti}}, \bibinfo {author}
  {\bibfnamefont {S.}~\bibnamefont {Romeo}}, \bibinfo {author} {\bibfnamefont
  {J.}~\bibnamefont {Scifo}}, \bibinfo {author} {\bibfnamefont
  {V.}~\bibnamefont {Shpakov}},  \emph {et~al.},\ }\href@noop {} {\bibfield
  {journal} {\bibinfo  {journal} {Optics Letters}\ }\textbf {\bibinfo {volume}
  {46}},\ \bibinfo {pages} {2844} (\bibinfo {year} {2021})}\BibitemShut
  {NoStop}%
\bibitem [{CLE()}]{CLEARgun}%
  \BibitemOpen
  \href@noop {} {\enquote {\bibinfo {title} {{RF gun for the CLIC
  accelerator}},}\ }\bibinfo {howpublished}
  {\url{http://cds.cern.ch/record/2753472?ln=en}}\BibitemShut {NoStop}%
\bibitem [{\citenamefont {Alesini}\ \emph {et~al.}(2022)\citenamefont {Alesini}
  \emph {et~al.}}]{alesini:ipac2022-MOPOMS019}%
  \BibitemOpen
  \bibfield  {author} {\bibinfo {author} {\bibfnamefont {D.}~\bibnamefont
  {Alesini}} \emph {et~al.},\ }in\ \href {\doibase
  10.18429/JACoW-IPAC2022-MOPOMS019} {\emph {\bibinfo {booktitle} {Proc.
  IPAC'22}}},\ \bibinfo {series and number} {\bibinfo {series} {International
  Particle Accelerator Conference}\ No.~\bibinfo {number} {13}}\ (\bibinfo
  {publisher} {JACoW Publishing, Geneva, Switzerland},\ \bibinfo {year}
  {2022})\ pp.\ \bibinfo {pages} {671--674}\BibitemShut {NoStop}%
\bibitem [{ber()}]{bergoz}%
  \BibitemOpen
  \href {https://www.bergoz.com/} {\enquote {\bibinfo {title}
  {https://www.bergoz.com/},}\ }\BibitemShut {NoStop}%
\bibitem [{\citenamefont {Alesini}\ \emph {et~al.}()\citenamefont {Alesini}
  \emph {et~al.}}]{AlesiniPatent}%
  \BibitemOpen
  \bibfield  {author} {\bibinfo {author} {\bibfnamefont {D.}~\bibnamefont
  {Alesini}} \emph {et~al.},\ }\href@noop {} {\enquote {\bibinfo {title}
  {{International Patent Publication No. WO 2016/147118 A1, assigned to
  INFN}},}\ }\BibitemShut {NoStop}%
\bibitem [{hfs()}]{hfss_website}%
  \BibitemOpen
  \href {http://www.ansys.com/products/electronics/ansys-electronics-desktop}
  {\enquote {\bibinfo {title} {{ANSYS Electronics Desktop website}},}\
  }\BibitemShut {NoStop}%
\bibitem [{\citenamefont {Alesini}\ \emph {et~al.}(2015)\citenamefont
  {Alesini}, \citenamefont {Battisti}, \citenamefont {Ferrario}, \citenamefont
  {Foggetta}, \citenamefont {Lollo}, \citenamefont {Ficcadenti}, \citenamefont
  {Pettinacci}, \citenamefont {Custodio}, \citenamefont {Pirez}, \citenamefont
  {Musumeci} \emph {et~al.}}]{alesini2015new}%
  \BibitemOpen
  \bibfield  {author} {\bibinfo {author} {\bibfnamefont {D.}~\bibnamefont
  {Alesini}}, \bibinfo {author} {\bibfnamefont {A.}~\bibnamefont {Battisti}},
  \bibinfo {author} {\bibfnamefont {M.}~\bibnamefont {Ferrario}}, \bibinfo
  {author} {\bibfnamefont {L.}~\bibnamefont {Foggetta}}, \bibinfo {author}
  {\bibfnamefont {V.}~\bibnamefont {Lollo}}, \bibinfo {author} {\bibfnamefont
  {L.}~\bibnamefont {Ficcadenti}}, \bibinfo {author} {\bibfnamefont
  {V.}~\bibnamefont {Pettinacci}}, \bibinfo {author} {\bibfnamefont
  {S.}~\bibnamefont {Custodio}}, \bibinfo {author} {\bibfnamefont
  {E.}~\bibnamefont {Pirez}}, \bibinfo {author} {\bibfnamefont
  {P.}~\bibnamefont {Musumeci}},  \emph {et~al.},\ }\href@noop {} {\bibfield
  {journal} {\bibinfo  {journal} {Physical Review Special Topics-Accelerators
  and Beams}\ }\textbf {\bibinfo {volume} {18}},\ \bibinfo {pages} {092001}
  (\bibinfo {year} {2015})}\BibitemShut {NoStop}%
\bibitem [{\citenamefont {Alesini}\ \emph {et~al.}(2018)\citenamefont
  {Alesini}, \citenamefont {Battisti}, \citenamefont {Bellaveglia},
  \citenamefont {Cardelli}, \citenamefont {Falone}, \citenamefont {Gallo},
  \citenamefont {Lollo}, \citenamefont {Palmer}, \citenamefont {Pellegrino},
  \citenamefont {Piersanti} \emph {et~al.}}]{alesini2018design}%
  \BibitemOpen
  \bibfield  {author} {\bibinfo {author} {\bibfnamefont {D.}~\bibnamefont
  {Alesini}}, \bibinfo {author} {\bibfnamefont {A.}~\bibnamefont {Battisti}},
  \bibinfo {author} {\bibfnamefont {M.}~\bibnamefont {Bellaveglia}}, \bibinfo
  {author} {\bibfnamefont {F.}~\bibnamefont {Cardelli}}, \bibinfo {author}
  {\bibfnamefont {A.}~\bibnamefont {Falone}}, \bibinfo {author} {\bibfnamefont
  {A.}~\bibnamefont {Gallo}}, \bibinfo {author} {\bibfnamefont
  {V.}~\bibnamefont {Lollo}}, \bibinfo {author} {\bibfnamefont {D.~T.}\
  \bibnamefont {Palmer}}, \bibinfo {author} {\bibfnamefont {L.}~\bibnamefont
  {Pellegrino}}, \bibinfo {author} {\bibfnamefont {L.}~\bibnamefont
  {Piersanti}},  \emph {et~al.},\ }\href@noop {} {\bibfield  {journal}
  {\bibinfo  {journal} {Physical Review Accelerators and Beams}\ }\textbf
  {\bibinfo {volume} {21}},\ \bibinfo {pages} {112001} (\bibinfo {year}
  {2018})}\BibitemShut {NoStop}%
\bibitem [{\citenamefont {Grudiev}\ \emph {et~al.}(2009)\citenamefont
  {Grudiev}, \citenamefont {Calatroni},\ and\ \citenamefont
  {Wuensch}}]{Grudiev}%
  \BibitemOpen
  \bibfield  {author} {\bibinfo {author} {\bibfnamefont {A.}~\bibnamefont
  {Grudiev}}, \bibinfo {author} {\bibfnamefont {S.}~\bibnamefont {Calatroni}},
  \ and\ \bibinfo {author} {\bibfnamefont {W.}~\bibnamefont {Wuensch}},\ }\href
  {\doibase 10.1103/PhysRevSTAB.12.102001} {\bibfield  {journal} {\bibinfo
  {journal} {Phys. Rev. ST Accel. Beams}\ }\textbf {\bibinfo {volume} {12}},\
  \bibinfo {pages} {102001} (\bibinfo {year} {2009})}\BibitemShut {NoStop}%
\bibitem [{com()}]{comeb}%
  \BibitemOpen
  \href {http://www.comeb.it/} {\enquote {\bibinfo {title} {Comeb s.r.l.}}\
  }\BibitemShut {NoStop}%
\bibitem [{\citenamefont {Pompili}\ \emph {et~al.}(2016)\citenamefont
  {Pompili}, \citenamefont {Anania}, \citenamefont {Bellaveglia}, \citenamefont
  {Biagioni}, \citenamefont {Bisesto}, \citenamefont {Chiadroni}, \citenamefont
  {Cianchi}, \citenamefont {Croia}, \citenamefont {Curcio}, \citenamefont
  {Di~Giovenale} \emph {et~al.}}]{pompili2016beam}%
  \BibitemOpen
  \bibfield  {author} {\bibinfo {author} {\bibfnamefont {R.}~\bibnamefont
  {Pompili}}, \bibinfo {author} {\bibfnamefont {M.}~\bibnamefont {Anania}},
  \bibinfo {author} {\bibfnamefont {M.}~\bibnamefont {Bellaveglia}}, \bibinfo
  {author} {\bibfnamefont {A.}~\bibnamefont {Biagioni}}, \bibinfo {author}
  {\bibfnamefont {F.}~\bibnamefont {Bisesto}}, \bibinfo {author} {\bibfnamefont
  {E.}~\bibnamefont {Chiadroni}}, \bibinfo {author} {\bibfnamefont
  {A.}~\bibnamefont {Cianchi}}, \bibinfo {author} {\bibfnamefont
  {M.}~\bibnamefont {Croia}}, \bibinfo {author} {\bibfnamefont
  {A.}~\bibnamefont {Curcio}}, \bibinfo {author} {\bibfnamefont
  {D.}~\bibnamefont {Di~Giovenale}},  \emph {et~al.},\ }\href@noop {}
  {\bibfield  {journal} {\bibinfo  {journal} {Nuclear Instruments and Methods
  in Physics Research Section A: Accelerators, Spectrometers, Detectors and
  Associated Equipment}\ }\textbf {\bibinfo {volume} {829}},\ \bibinfo {pages}
  {17} (\bibinfo {year} {2016})}\BibitemShut {NoStop}%
\bibitem [{\citenamefont {Chiadroni}\ \emph {et~al.}(2017)\citenamefont
  {Chiadroni}, \citenamefont {Alesini}, \citenamefont {Anania}, \citenamefont
  {Bacci}, \citenamefont {Bellaveglia}, \citenamefont {Biagioni}, \citenamefont
  {Bisesto}, \citenamefont {Cardelli}, \citenamefont {Castorina}, \citenamefont
  {Cianchi} \emph {et~al.}}]{chiadroni2017beam}%
  \BibitemOpen
  \bibfield  {author} {\bibinfo {author} {\bibfnamefont {E.}~\bibnamefont
  {Chiadroni}}, \bibinfo {author} {\bibfnamefont {D.}~\bibnamefont {Alesini}},
  \bibinfo {author} {\bibfnamefont {M.}~\bibnamefont {Anania}}, \bibinfo
  {author} {\bibfnamefont {A.}~\bibnamefont {Bacci}}, \bibinfo {author}
  {\bibfnamefont {M.}~\bibnamefont {Bellaveglia}}, \bibinfo {author}
  {\bibfnamefont {A.}~\bibnamefont {Biagioni}}, \bibinfo {author}
  {\bibfnamefont {F.}~\bibnamefont {Bisesto}}, \bibinfo {author} {\bibfnamefont
  {F.}~\bibnamefont {Cardelli}}, \bibinfo {author} {\bibfnamefont
  {G.}~\bibnamefont {Castorina}}, \bibinfo {author} {\bibfnamefont
  {A.}~\bibnamefont {Cianchi}},  \emph {et~al.},\ }\href@noop {} {\bibfield
  {journal} {\bibinfo  {journal} {Nuclear Instruments and Methods in Physics
  Research Section A: Accelerators, Spectrometers, Detectors and Associated
  Equipment}\ } (\bibinfo {year} {2017})}\BibitemShut {NoStop}%
\bibitem [{\citenamefont {Bettoni}\ \emph {et~al.}(2018)\citenamefont
  {Bettoni}, \citenamefont {Craievich}, \citenamefont {Pedrozzi}, \citenamefont
  {Schaer},\ and\ \citenamefont {Stingelin}}]{bettoni2018low}%
  \BibitemOpen
  \bibfield  {author} {\bibinfo {author} {\bibfnamefont {S.}~\bibnamefont
  {Bettoni}}, \bibinfo {author} {\bibfnamefont {P.}~\bibnamefont {Craievich}},
  \bibinfo {author} {\bibfnamefont {M.}~\bibnamefont {Pedrozzi}}, \bibinfo
  {author} {\bibfnamefont {M.}~\bibnamefont {Schaer}}, \ and\ \bibinfo {author}
  {\bibfnamefont {L.}~\bibnamefont {Stingelin}},\ }\href@noop {} {\bibfield
  {journal} {\bibinfo  {journal} {Physical Review Accelerators and Beams}\
  }\textbf {\bibinfo {volume} {21}},\ \bibinfo {pages} {023401} (\bibinfo
  {year} {2018})}\BibitemShut {NoStop}%
\bibitem [{\citenamefont {Cardelli}\ \emph {et~al.}(2022)\citenamefont
  {Cardelli} \emph {et~al.}}]{cardelli:ipac2022-mopoms020}%
  \BibitemOpen
  \bibfield  {author} {\bibinfo {author} {\bibfnamefont {F.}~\bibnamefont
  {Cardelli}} \emph {et~al.},\ }in\ \href {\doibase
  10.18429/JACoW-IPAC2022-MOPOMS020} {\emph {\bibinfo {booktitle} {Proc.
  IPAC'22}}},\ \bibinfo {series and number} {\bibinfo {series} {International
  Particle Accelerator Conference}\ No.~\bibinfo {number} {13}}\ (\bibinfo
  {publisher} {JACoW Publishing, Geneva, Switzerland},\ \bibinfo {year}
  {2022})\ pp.\ \bibinfo {pages} {675--678}\BibitemShut {NoStop}%
\bibitem [{\citenamefont {Pompili}\ \emph {et~al.}(2020)\citenamefont
  {Pompili}, \citenamefont {Alesini}, \citenamefont {Anania}, \citenamefont
  {Behtouei}, \citenamefont {Bellaveglia}, \citenamefont {Biagioni},
  \citenamefont {Bisesto}, \citenamefont {Cesarini}, \citenamefont {Chiadroni},
  \citenamefont {Cianchi} \emph {et~al.}}]{pompili2020energy}%
  \BibitemOpen
  \bibfield  {author} {\bibinfo {author} {\bibfnamefont {R.}~\bibnamefont
  {Pompili}}, \bibinfo {author} {\bibfnamefont {D.}~\bibnamefont {Alesini}},
  \bibinfo {author} {\bibfnamefont {M.}~\bibnamefont {Anania}}, \bibinfo
  {author} {\bibfnamefont {M.}~\bibnamefont {Behtouei}}, \bibinfo {author}
  {\bibfnamefont {M.}~\bibnamefont {Bellaveglia}}, \bibinfo {author}
  {\bibfnamefont {A.}~\bibnamefont {Biagioni}}, \bibinfo {author}
  {\bibfnamefont {F.}~\bibnamefont {Bisesto}}, \bibinfo {author} {\bibfnamefont
  {M.}~\bibnamefont {Cesarini}}, \bibinfo {author} {\bibfnamefont
  {E.}~\bibnamefont {Chiadroni}}, \bibinfo {author} {\bibfnamefont
  {A.}~\bibnamefont {Cianchi}},  \emph {et~al.},\ }\href@noop {} {\bibfield
  {journal} {\bibinfo  {journal} {Nature Physics}\ ,\ \bibinfo {pages} {1}}
  (\bibinfo {year} {2020})}\BibitemShut {NoStop}%
\bibitem [{\citenamefont {Pompili}\ \emph {et~al.}(2022)\citenamefont
  {Pompili}, \citenamefont {Alesini}, \citenamefont {Anania}, \citenamefont
  {Arjmand}, \citenamefont {Behtouei}, \citenamefont {Bellaveglia},
  \citenamefont {Biagioni}, \citenamefont {Buonomo}, \citenamefont {Cardelli},
  \citenamefont {Carpanese} \emph {et~al.}}]{pompili2022free}%
  \BibitemOpen
  \bibfield  {author} {\bibinfo {author} {\bibfnamefont {R.}~\bibnamefont
  {Pompili}}, \bibinfo {author} {\bibfnamefont {D.}~\bibnamefont {Alesini}},
  \bibinfo {author} {\bibfnamefont {M.}~\bibnamefont {Anania}}, \bibinfo
  {author} {\bibfnamefont {S.}~\bibnamefont {Arjmand}}, \bibinfo {author}
  {\bibfnamefont {M.}~\bibnamefont {Behtouei}}, \bibinfo {author}
  {\bibfnamefont {M.}~\bibnamefont {Bellaveglia}}, \bibinfo {author}
  {\bibfnamefont {A.}~\bibnamefont {Biagioni}}, \bibinfo {author}
  {\bibfnamefont {B.}~\bibnamefont {Buonomo}}, \bibinfo {author} {\bibfnamefont
  {F.}~\bibnamefont {Cardelli}}, \bibinfo {author} {\bibfnamefont
  {M.}~\bibnamefont {Carpanese}},  \emph {et~al.},\ }\href@noop {} {\bibfield
  {journal} {\bibinfo  {journal} {Nature}\ }\textbf {\bibinfo {volume} {605}},\
  \bibinfo {pages} {659} (\bibinfo {year} {2022})}\BibitemShut {NoStop}%
\bibitem [{\citenamefont {L.M.Young}()}]{TStep2}%
  \BibitemOpen
  \bibfield  {author} {\bibinfo {author} {\bibnamefont {L.M.Young}},\ }\href
  {https://lmytechnology.com/} {\enquote {\bibinfo {title} {Tstep: an electron
  linac design code},}\ }\BibitemShut {NoStop}%
\bibitem [{gpt()}]{gpt_web}%
  \BibitemOpen
  \href@noop {} {\enquote {\bibinfo {title} {{Pulsar Physics}},}\ }\bibinfo
  {howpublished} {\url{http://www.pulsar.nl/gpt}}\BibitemShut {NoStop}%
\bibitem [{\citenamefont {Di~Mitri}\ \emph {et~al.}(2013)\citenamefont
  {Di~Mitri}, \citenamefont {Allaria}, \citenamefont {Castronovo},
  \citenamefont {Cornacchia}, \citenamefont {Fawley}, \citenamefont
  {Fr{\"o}hlich}, \citenamefont {Karantzoulis}, \citenamefont {Penco},
  \citenamefont {Serpico}, \citenamefont {Spezzani} \emph
  {et~al.}}]{di2013emittance}%
  \BibitemOpen
  \bibfield  {author} {\bibinfo {author} {\bibfnamefont {S.}~\bibnamefont
  {Di~Mitri}}, \bibinfo {author} {\bibfnamefont {E.}~\bibnamefont {Allaria}},
  \bibinfo {author} {\bibfnamefont {D.}~\bibnamefont {Castronovo}}, \bibinfo
  {author} {\bibfnamefont {M.}~\bibnamefont {Cornacchia}}, \bibinfo {author}
  {\bibfnamefont {W.}~\bibnamefont {Fawley}}, \bibinfo {author} {\bibfnamefont
  {L.}~\bibnamefont {Fr{\"o}hlich}}, \bibinfo {author} {\bibfnamefont
  {E.}~\bibnamefont {Karantzoulis}}, \bibinfo {author} {\bibfnamefont
  {G.}~\bibnamefont {Penco}}, \bibinfo {author} {\bibfnamefont
  {C.}~\bibnamefont {Serpico}}, \bibinfo {author} {\bibfnamefont
  {C.}~\bibnamefont {Spezzani}},  \emph {et~al.},\ }in\ \href@noop {} {\emph
  {\bibinfo {booktitle} {Proc. FEL 2013, NY}}}\ (\bibinfo {year} {2013})\ pp.\
  \bibinfo {pages} {6--11}\BibitemShut {NoStop}%
\bibitem [{\citenamefont {Palmer}\ \emph {et~al.}(1997)\citenamefont {Palmer},
  \citenamefont {Wang}, \citenamefont {Miller}, \citenamefont {Babzien},
  \citenamefont {Ben-Zvi}, \citenamefont {Pellegrini}, \citenamefont {Sheehan},
  \citenamefont {Skaritka}, \citenamefont {Winick}, \citenamefont {Woodle}
  \emph {et~al.}}]{palmer1997initial}%
  \BibitemOpen
  \bibfield  {author} {\bibinfo {author} {\bibfnamefont {D.}~\bibnamefont
  {Palmer}}, \bibinfo {author} {\bibfnamefont {X.}~\bibnamefont {Wang}},
  \bibinfo {author} {\bibfnamefont {R.}~\bibnamefont {Miller}}, \bibinfo
  {author} {\bibfnamefont {M.}~\bibnamefont {Babzien}}, \bibinfo {author}
  {\bibfnamefont {I.}~\bibnamefont {Ben-Zvi}}, \bibinfo {author} {\bibfnamefont
  {C.}~\bibnamefont {Pellegrini}}, \bibinfo {author} {\bibfnamefont
  {J.}~\bibnamefont {Sheehan}}, \bibinfo {author} {\bibfnamefont
  {J.}~\bibnamefont {Skaritka}}, \bibinfo {author} {\bibfnamefont
  {H.}~\bibnamefont {Winick}}, \bibinfo {author} {\bibfnamefont
  {M.}~\bibnamefont {Woodle}},  \emph {et~al.},\ }in\ \href@noop {} {\emph
  {\bibinfo {booktitle} {AIP Conference Proceedings}}},\ Vol.\ \bibinfo
  {volume} {398}\ (\bibinfo {organization} {American Institute of Physics},\
  \bibinfo {year} {1997})\ pp.\ \bibinfo {pages} {695--704}\BibitemShut
  {NoStop}%
\bibitem [{\citenamefont {Zheng}\ \emph {et~al.}(2016)\citenamefont {Zheng},
  \citenamefont {Du}, \citenamefont {Zhang}, \citenamefont {Qian},
  \citenamefont {Yan}, \citenamefont {Shi}, \citenamefont {Zhang},
  \citenamefont {Zhou}, \citenamefont {Wu}, \citenamefont {Su} \emph
  {et~al.}}]{zheng2016development}%
  \BibitemOpen
  \bibfield  {author} {\bibinfo {author} {\bibfnamefont {L.}~\bibnamefont
  {Zheng}}, \bibinfo {author} {\bibfnamefont {Y.}~\bibnamefont {Du}}, \bibinfo
  {author} {\bibfnamefont {Z.}~\bibnamefont {Zhang}}, \bibinfo {author}
  {\bibfnamefont {H.}~\bibnamefont {Qian}}, \bibinfo {author} {\bibfnamefont
  {L.}~\bibnamefont {Yan}}, \bibinfo {author} {\bibfnamefont {J.}~\bibnamefont
  {Shi}}, \bibinfo {author} {\bibfnamefont {Z.}~\bibnamefont {Zhang}}, \bibinfo
  {author} {\bibfnamefont {Z.}~\bibnamefont {Zhou}}, \bibinfo {author}
  {\bibfnamefont {X.}~\bibnamefont {Wu}}, \bibinfo {author} {\bibfnamefont
  {X.}~\bibnamefont {Su}},  \emph {et~al.},\ }\href@noop {} {\bibfield
  {journal} {\bibinfo  {journal} {Nuclear Instruments and Methods in Physics
  Research Section A: Accelerators, Spectrometers, Detectors and Associated
  Equipment}\ }\textbf {\bibinfo {volume} {834}},\ \bibinfo {pages} {98}
  (\bibinfo {year} {2016})}\BibitemShut {NoStop}%
\bibitem [{\citenamefont {Divall}\ \emph {et~al.}(2015)\citenamefont {Divall},
  \citenamefont {Prat}, \citenamefont {Bettoni}, \citenamefont {Vicario},
  \citenamefont {Trisorio}, \citenamefont {Schietinger},\ and\ \citenamefont
  {Hauri}}]{divall2015intrinsic}%
  \BibitemOpen
  \bibfield  {author} {\bibinfo {author} {\bibfnamefont {M.~C.}\ \bibnamefont
  {Divall}}, \bibinfo {author} {\bibfnamefont {E.}~\bibnamefont {Prat}},
  \bibinfo {author} {\bibfnamefont {S.}~\bibnamefont {Bettoni}}, \bibinfo
  {author} {\bibfnamefont {C.}~\bibnamefont {Vicario}}, \bibinfo {author}
  {\bibfnamefont {A.}~\bibnamefont {Trisorio}}, \bibinfo {author}
  {\bibfnamefont {T.}~\bibnamefont {Schietinger}}, \ and\ \bibinfo {author}
  {\bibfnamefont {C.~P.}\ \bibnamefont {Hauri}},\ }\href@noop {} {\bibfield
  {journal} {\bibinfo  {journal} {Physical Review Special Topics-Accelerators
  and Beams}\ }\textbf {\bibinfo {volume} {18}},\ \bibinfo {pages} {033401}
  (\bibinfo {year} {2015})}\BibitemShut {NoStop}%
\end{thebibliography}%





\end{document}